\documentstyle[preprint,aps,rotate,epsfig]{revtex}
\tighten
\begin{document}
\title {\bf Nuclear equation of state at high density
and the properties of neutron stars }
\author{P. K. Sahu\\
Division of Physics, Graduate School of Science\\
Hokkaido University, Sapporo 060-0810\\
JAPAN}

\date{\today}

\maketitle
\begin{abstract}
We discuss the relativistic nuclear equation of state (EOS) using a
relativistic transport model in heavy-ion collisions. From 
the baryon flow for $Au + Au$ systems at SIS to AGS energies and above 
we find that the strength of the vector potential has to be reduced 
moderately at high density or at high relative momenta to describe the 
flow data at 1-10 A GeV. We use the same dynamical model
to calculate the nuclear EOS and then employ this to calculate the
gross structure of the neutron star considering the core 
to be composed of neutrons with an admixture of protons, 
electrons, muons, sigmas and lambdas at zero temperature. We then discuss
these gross properties of neutron stars such as maximum mass and radius 
in contrast to the observational values.
\end{abstract}

{\noindent PACS number(s): 26.60.+c, 21.65.+f}

\newpage
\section{Introduction}
The nuclear equation of state (EOS) at high density is still an unresolved 
issue though many theoretical and experimental efforts have been made
in the last two decades to address this question in a more systematic way. 
Theoretically, specially in astrophysics, the density of the core inside the
compact objects like neutron star is greater than normal nuclear matter 
density, composed of many non-strange and strange degrees of freedom.
One of the most important characteristic feature of a neutron star is its 
maximum allowed mass.
The determination of maximum mass and radius of neutron star are 
dominated by the interactions between particles at high density 
and its EOS.
There are many models available in the literature to deal with
maximum masses of neutron stars.
These are relativistic and non- relativistic approaches. 
The non- relativistic models \cite{akma98,wiri88} based on the potential
approach describe the nuclear structure for light nuclei. 
However, the relativistic models \cite{sero86,lang91,glen92,ghos95,sahu93} 
constructed from Lagrangian approach explain the nuclear structure 
data for heavy nuclei without violating the properties of nuclear matter
at saturation density. 
In both conventional approaches in the neutron star matter, the 
estimated maximum masses of neutron star are above $2M_{\odot}$.
Recently from several calculations, it has been pointed out 
\cite{heis99,balb97,shen98} that the nuclear 
EOS should be soft at high density.
This is due to fact that all the measured  neutron 
star masses are less than $2M{\odot}$ \cite{thor99}.
Various scenarios including the reduced strength of vector field, the
presence of hyperons and possibility of kaon condensation, have been 
proposed to be soften the EOS.

\par
Regarding the composition of neutron star matter, there are 
calculations \cite{brow94}, which include kaons as the strange 
particles along with neutrons and protons e.g., the possibility of 
kaon condensation.
Also there are models \cite{glen92,ghos95} in 
the neutron star matter where the composition of particles are sigmas and 
lambdas as strange particles besides neutrons, protons and electrons as 
non- strange particles.
Both these proposed models in the neutron star matter lead to 
a soft EOS at high density.
In this paper we consider the existence of hyperons in the neutron star
matter with the recent compiled information of nuclear interaction 
from heavy-ion collisions.
\par
Experimentally, the nuclear EOS is very important to understand the non- 
equilibrium complicated heavy- ion collisions data at very high energies.
Very recently \cite{liu98,pink99}, the heavy- ion collisions data such as 
the sideward and elliptic flow have been measured at AGS energies. 
The sideward flow data are mainly determined by the nature of the nuclear 
force in the nuclear EOS. 
Moreover, the nuclear EOS can be understood better from
the elliptic flow than sideward flow, because the elliptical
flow plays less uncertainties role in the opposing stream of matter moving
past each other within the reaction plane in the heavy- ion collisions.
Recently, the beam energy dependence of flow data \cite{liu98,pink99} 
indicate that the nuclear EOS is rather soft to lead a possible phase
transition to quark gluon plasma at high density and hence the strength 
of the repulsive vector potential must be low to describe these data in 
the heavy- ion collisions.
\par
In the present discussion, we use an extended version of relativistic mean 
field model \cite{sahu98} including the momenta dependent forces, which are 
taken into account phenomenologically in the relativistic transport model in 
heavy- ion collisions. 
We calculate the nuclear EOS by using the same dynamic momentum dependence 
constraints in the nuclear potentials and then employ to the neutron 
star structure calculations.
The aim of this paper is to derive the nuclear force from the 
heavy-ion collisions data e.g., from nucleon flow data and then 
to study this force on the gross structure of the neutron star by 
giving less importance to the composition of neutron star matter.
As far as the strange particles are concerned, we take minimum strange 
particles ($\Sigma$ and $\Lambda$) in the neutron star matter calculation
at high densities.
\par
The paper is organized as follows: In section 2 we briefly describe the 
relativistic nuclear EOS and its derivation from heavy- ion reactions. 
In section 3 we employ the same nuclear EOS to the neutron star structure 
with the systematic results. 
The conclusion and summary are presented in section 4. 
\section{Relativistic nuclear equation of state}
The relativistic mean- field theory is very successful model in the 
relativistic transport model in heavy- ion collisions as well as 
in nuclear structure physics.
Originally, Walecka \cite{sero86} had proposed the relativistic mean- field 
model and later the modified version of this has been used widely to 
calculate nuclear structure and nuclear matter properties. 
The extended version of the Walecka model, so called non- linear relativistic
mean- field model \cite{lang91,ghos95} has the interaction of Dirac nucleons 
with scalar and 
vector mesons as well as non- linear self- interaction of the scalar field.
The extra non- linear self- interaction scalar field helps to get the empirical 
values of bulk properties of nuclear matter at saturation density, e.g., the 
nuclear incompressibility and the value of effective nucleon mass in the 
desirable range.
The physics behind this phenomenological successful model is that the 
nucleon- nucleon interaction in the mean- field theory contains strong
attractive Lorentz scalar and repulsive Lorentz vector components, which 
almost cancel for low momenta, but produce a strong spin- orbit force 
consistent with the observed single- particle spectra.
In the original Walecka model \cite{sero86}, the vector potential increases linearly with
density, whereas the scalar potential changes non- linearly. 
This is because the vector and scalar potentials have linear and non- linear
function of density respectively. 
However, from the heavy- ion collisions data, we find that the vector 
potential also should have non- linear function of baryon density, i.e., the
strength of the vector potential should be low at high density \cite{sahu98} 
compared to the original model \cite{sero86}. 
Recently, this fact has been taken by adding the non-
linear vector meson terms in the original Lagrangian density and applied to
the nuclear matter, neutron star matter \cite{shen98} and nuclear 
structure \cite{suga94} calculations.
In our calculation, we take the non- linear effect in the vector meson
with density by employing the phenomenological momentum dependent cut- off
to the vector potential term.
We adopt this method keeping in mind to describe the heavy- ion 
reactions data at high energies, which generates the nuclear matter
like situation in the laboratory.
We recall that the mean-field energy density 
for nuclear matter in the relativistic mean-field model can be written 
as \cite{lang91}
\begin{eqnarray}
\varepsilon(m^*,n_b)
&=&g_v V_0 n_b -\frac{1}{2}m_v^2 V_0^2 
+ \frac{m_s^2}{2 g_s^2}(m-m^*)^2 
+ \frac{B}{3 g_s^3}(m-m^*)^3     \nonumber \\
&+& \frac{C}{4 g_s^4}(m-m^*)^4 + \gamma\int^{k_f}_{0} \frac{d^3p}{(2\pi)^3}
\sqrt{(p^2+m^*)},  
\label{energy}
\end{eqnarray}
where $m^*=m-g_s S_0$ is the effective nucleon mass, $n_b$ is
the baryon density and the spin and isospin degeneracy is $\gamma=4$.
$S_0$ and $V_0$ are the scalar and vector fields with mass $m_s$ and
$m_{v}$, which couple to nucleons with coupling constants $g_s$ and $g_v$, 
respectively. 
$B$ and $C$ are constant parameters describing the scalar 
self-interactions field  and $p$ is the
nucleon momentum integrated up to the Fermi momentum $k_f$.
In (1), the vector and scalar potentials depend on density, 
however, the vector potential increases linearly with density ($n_B$).
The parameters $g_v$, $g_s$, $B$ and $C$ in (1) are determined by fitting
the saturation density, binding energy, effective nucleon mass and
the compression modulus at nuclear matter density (cf. NL3 parameters set 
from Table I in Ref. \cite{lang91}).
\par
In our present calculation, we have extended (1) to include a non-linear
dependence of the vector potential on the baryon density by implementing
the momentum ($p$) dependent form factor at the vertices and can be written 
as \cite{sahu98}
\begin{equation}
V_0(p)=V_0 \frac{{ p}^2-\Lambda_{v1}^2}{{ p}^2+\Lambda_{v2}^2},
\end{equation}
where the cut- off parameters $\Lambda_{v1}=0.37$ GeV, $\Lambda_{v2}=0.9$ 
GeV and $V_0$ is vector potential.
For completeness, we incorporate the momentum dependent form factor at the 
scalar vertices in the form given as \cite{sahu98}
\begin{equation}
V_s(p)=V_s \frac{{ p}^2-\Lambda_{s1}^2}{{ p}^2+\Lambda_{s2}^2},
\end{equation}
where the cut- off parameters $\Lambda_{s1}=0.71$ GeV, $\Lambda_{s2}=1.0$ 
GeV and $V_s$ is scalar potential.
The choice of these form factors are similar to that used in effective
meson-exchange interactions for nucleon-nucleon scattering \cite{mach87}
and later this strategy was used in relativistic approach for nucleus-nucleus
collisions from SIS to SPS energies\cite{eheh96}.
The values of cut-off parameters 
in vector and scalar vertices are chosen to describe properly the 
Schr{\"o}dinger-equivalent potential until 1 GeV and the flow data
at AGS energies.
These cut-off parameters are not unique for various type of equation 
of states to fit Schr{\"o}dinger-equivalent potential until 1 GeV and 
the flow data at AGS energies simultaneously. 
We note that the form factor eq.(2) will make the vector interaction weak
at high baryon density and at high energies in heavy- ion collisions.
At these energies, it has also been observed that the strength of repulsive 
vector potential should be reduced considerably at high density or at
high relative momenta to describe the flow data.
Theoretically, it is important to understand the decrease of vector
coupling at high density.
Contrast to heavy-ion reactions, in this line some works have been
performed \cite{song97} and more to be required in 
details \cite{sahup1}.
\par
We show in Fig. 1, the scalar and vector potential energies as a function of
baryon density. 
The solid lines (NLE curves) are associated with the momentum dependent 
form factors given in eqs(2,3) that describe the flow data best from
SIS to AGS energies.
The dashed lines (NL3 curves) are without momentum dependent potentials.
The vector part for NLE is substantially low at high baryon density
as that of NL3 parameter set.
At high density the reduction of vector potential is more significant
than the
scalar potential for NLE curves.
Therefore, the net effect of changed potentials is vector potential 
due to substantial reduction of vector part at high baryon density. 
For example, at $\rho=8\rho_0$, the values of vector part and scalar part are
1250 MeV and  -511 MeV respectively for NLE, where the value of 
vector part is 1740 MeV and scalar part is -735 MeV for NL3.
So, the net reduction is dominated by the vector potential in NLE model.
The corresponding EOS versus baryon density are shown in Fig. 2
for extended momentum dependent
model (NLE) as well as the original non- linear model (NL3). 
NLE has the  momentum dependent form factor in the vector and scalar
potentials.
The other nuclear EOS has been discussed more details in the Ref. \cite{ghos95}
by varying the nuclear incompressibility from low (soft) 250 MeV to high 
(stiff) 350 MeV values. 
We do not elaborate on that issue here, because 
we would like to emphasize more on the momentum dependent force in the 
nuclear EOS along the line of heavy-ion reaction data.
We see in Fig. 2 that NLE nuclear EOS is softer than NL3 at density
$\ge 7 \rho_0$ and is slightly stiffer at density $\le 7\rho_0$.
The incompressibility is close ($\sim$ 380 MeV) to NL3 value at saturation 
density.
In the next section, we would like to implement this model in the
neutron star matter, where the core density is in the range of
$> 5-8 \rho_0$.
So, in the present model, the stiffness of equation of state changes
around that density, due to the main contribution coming from the
reduced vector potential.
However, in the heavy- ion flow calculation at AGS energies, the 
stiffness of equation of state not only comes from the net reduction of
vector potential but also from the transition from hadron to 
string degrees of freedom as discussed in our recent work \cite{sahu98}.
It has been pointed out recently\cite{frim97} from the 
simulation calculation that
one might even reach $10\rho_0$, although only for a very short time
of a few fm/c at energy range between the AGS and the SPS energies.
Hence at AGS energy range, the baryon density is expected to reach more 
than $>5\rho_0$. 
\par
Recently, the elliptic flow and the sideward flow have been studied
theoretically with increasing beam energy by various type of equation 
of state and the possible signature of phase transition\cite{liu00}.
More precisely, the beam energy dependence of the observed elliptic
flow has been interpreted such a possible phase transition.
The reason is that a simulation model including different kinds of 
equation of state is consistent with a softening of the equation of 
state.
This softening of equation of state can be realized in many ways, 
for example: (i) by reducing the strong repulsive force in the 
equation of state with help of momentum dependent form
factor and fitting it with Schr{\"o}dinger-equivalent potential   
and (ii) by implementing transition from hadronic to
string degrees of freedom with beam energies in the simulation 
model\cite{sahu98}.
In our calculation, we implement the former one, where the thermodynamic 
pressure in the extended model NLE is lower as compared to the NL3 model 
due to less repulsive force at AGS energy regime. 
We thus get reduced repulsive force because of the strong cut-off 
parameters eq. (2) in the vector potential. 
Also this cut-off makes the vector potential non-linear
function of baryon density. 
\section{Neutron star matter and properties of neutron star}
\subsection{Neutron star matter}
The core of the neutron star plays a significant role to determine the 
gross structural properties like maximum mass and radius of the neutron
star.
The density of the core inside the neutron star is greater than the 
normal nuclear matter density and hence the nuclear interactions are 
important to construct the neutron star matter EOS around that density.
Moreover, in such a high density, the strange particles are expected to be 
present along with usual neutron matters like neutrons, protons and electrons.
So, in our neutron star matter calculation we assume that the core of the 
neutron star matter is composed of neutrons with an admixture of protons,
electrons, muons and hyperons ($\Lambda$ and $\Sigma^-$) \cite{ghos95}.
The concentrations of each particle can be determined by using the condition of 
equilibrium under the weak interactions (assuming that neutrinos are not
degenerate) and the electric charge neutrality:
\begin{eqnarray}
\mu_p=\mu_n-\mu_e,~~ \mu_{\Lambda}=\mu_n,\nonumber \\
 \mu_{\Sigma^-}=\mu_n+\mu_e, ~~
\mu_{\mu}=\mu_e ; \nonumber \\
n_p=n_e+n_{\mu}+n_{\Sigma^-}.
\end{eqnarray}
In addition, the total baryon density is $n_B=n_n+n_p+n_{\Lambda}+n_{\Sigma^-}$
and the baryon chemical potential is $\mu_B=\mu_n$, where $n_i$ and 
$\mu_i$ stand for number density and chemical potential for $i-$th particle 
respectively.
\par
Since the nuclear force is known to favor isospin symmetry and the symmetry
energy arising solely from the Fermi energy is known to be inadequate to account
for the empirical value of the symmetry energy ($\sim$ 32 MeV), we include
the interaction due to isospin triplet $\rho$- meson in the relativistic
non- linear mean field model for the purpose of describing the neutron- rich 
matter \cite{sahu93}.
It is noted that the $\rho$- meson will contribute a term $=\frac{g_{\rho}^2}
{8 m_{\rho}^2}(n_p-n_n)^2$ to the energy density and pressure.
We fix the coupling constant $g_{\rho}$ by requiring that the symmetric energy
coefficient correspond to the empirical value 32 MeV.
Then the neutron star matter EOS is calculated from energy density 
$\varepsilon$ and pressure $P$ are given as follow \cite{ghos95}:
\begin{eqnarray}
\varepsilon=\frac{1}{2}m_v^2 V_0^2 +\frac{1}{2}m_{\rho}^2 {\rho_0}^2 
+\frac{1}{2}m_s^2 S_0^2 +\frac{B}{3}S_0^3
\nonumber\\
 +\frac{C}{4}S_0^4 +\sum_i \varepsilon_{FG} 
+\sum_l \varepsilon_{FG}
\nonumber\\
P=\frac{1}{2}m_v^2 V_0^2 +\frac{1}{2}m_{\rho}^2 {\rho_0}^2 
-\frac{1}{2}m_s^2 S_0^2 -\frac{B}{3}S_0^3
\nonumber\\
 -\frac{C}{4}S_0^4 +\sum_i P_{FG} 
+\sum_l P_{FG}
\end{eqnarray}
where $\rho_0$ is the third component in isospin space. 
In the above equations $\varepsilon_{FG}$ and $P_{FG}$ are the relativistic
non- interacting energy density and pressure of the baryons ($i$) and leptons
($l$) respectively.
\par
The three coupling constant parameters of hyperon- meson interaction are not 
well known.
Therefore, we fix the ratio of hyperon-meson and nucleon-meson couplings
for $\sigma$, $\omega$ and $\rho$ mesons respectively (i) by choosing very 
close to the quark counting rule \cite{ghos95} e.g., the 
potentials seen by $\Lambda$ and $\Sigma$ in nuclear matter are $\sim -30$ MeV
\cite{chri89} and (ii) assuming the attractive potential seen by $\Lambda$ 
and repulsive potential seen by $\Sigma$ to be
$\sim$ -30 MeV \cite{chri89} and $\sim$ +10 MeV \cite{balb97,yama99,dabr99} 
respectively, at nuclear matter density.
The analysis of various experimental data on 
hypernuclei \cite{chri89,yama99,dabr99,hara91} 
suggest that the strength of $\Sigma$ potential may be either repulsive or 
attractive at nuclear matter density.
This point will be cleared further after analysis of more hypernuclei
data in near future and the general discussions are given in 
recent Ref. \cite{rijk99}. 
Due to this fact, we consider the two possibility of strength on $\Sigma$
potential as discussed above. 
\par
Taking all these parameters into the equations (4), we show the 
concentration of particles ($x_i=n_i/n_B$, i=$p$, $\Sigma$ and $\Lambda$) 
versus baryon density for NLE1, NLE2 and NL3 models in Fig. 3. 
We display $p$, $\Sigma$ and $\Lambda$ particles in this figure due
to practical importance in neutron stars, for example, $p$ fraction plays 
role for cooling process of neutron stars and the order of appearance of 
strange particles with density may influence the EOS of the neutron star 
matter.
In NLE1 and NL3 models, the potentials for $\Lambda$ and $\Sigma$ are taken
to be equal to $\sim$ -30 MeV, where the potentials for $\Lambda$ and 
$\Sigma$ are chosen to be $\sim$ -30 MeV and $\sim$ +10 MeV respectively in 
NLE2 model. 
However, the momentum dependent cut-off to the vector potential are 
incorporated in both NLE1 and NLE2 models.
We notice in Fig. 3 that the concentration of particles like 
$\Sigma^-$ and $\Lambda$ start appearing after 2 times 
nuclear matter density for all models. 
In NLE1 and NL3 models, the order of appearance of strange particles 
(first $\Sigma$ and then $\Lambda$) are same due to equal strength of 
potential felt by strange particles.
Where the situation is quite different in case of NLE2 model, here $\Lambda$
appears first around $>2.5$ time nuclear matter density and $\Sigma^-$ starts
coming much later \cite{balb97} around $>3.5$ times nuclear matter density.
This is due to fact that $\Sigma$ sees extra strength +40 MeV potential than
$\Lambda$ potential, which is repulsive.
In both NLE1 and NLE2 models, the strange particles start coming slightly
later than NL3, due to reduction of vector potential by momentum dependent
cut-off as given in equation (2).
However, the change of proton concentration is not very significant with 
density for all models, except the slight decreasing tendency at high 
density was shown by NLE2.
At around 1.5 times nuclear matter density, the value of protons 
concentration crosses the threshold value 0.11 (horizontal line in Fig. 3), 
which shows that the direct URCA process is possible to lead for cooling 
of the neutron stars in all models \cite{peth92}.
\subsection{Maximum mass and radius of neutron star}
\par
The gross structure of the neutron stars such as mass and radius are calculated
from the equations that describe the hydrostatic equilibrium of degenerate
stars without rotation in general relativity is called Tolman-Oppenheimer-
Volkoff (TOV) equations \cite{sahu93}.
From the dynamics and transport properties of pulsars, the additional 
structure parameters of neutron stars like the moment of inertia $I$ and
the surface red shift $z=\frac{1} {\sqrt{1-2GM/Rc^2}} -1$ are important 
and are given more elaborately in Ref. \cite{sahu93}.
\par
We solve the TOV equations by constructing the EOS for the entire density
region starting from the higher density at the center to the surface density.
The composite EOS for the entire neutron star density span was constructed 
by joining the NLE and NL3 neutron star matter EOS to that EOS of the 
density range (i) $10^{14}$ to $5 \times 10^{10}~g~ cm^{-3}$ \cite{negl73}, (ii)
$5 \times 10^{10}~g ~cm^{-3}$ to $10^3~g ~cm^{-3}$ \cite{baym71}  and (iii) 
less than $10^3~g ~cm^{-3}$ \cite{feyn49}.
The composite neutron star matter EOS are plotted in Fig. 4 for NLE1, NLE2 
and NL3 models, which are used to calculate the neutron star structures 
as discussed above.
From Fig. 4 we find that the pressure is low at high density for NLE1, NLE2 
EOS and hence the soft EOS compare to NL3 EOS. 
If we look at the Fig. 3, the order of appearance of $\Lambda$ particles 
with density reflect in the same order of the nature of EOS.
That is, NL3 is stiffer than NLE1 and NLE2 EOS, because the momentum
dependent form factor in later two models has reduced the vector 
potential at high density. 
So, NLE2 is similar to NLE1 EOS, except a slight stiffer than NLE1 due to
the strong repulsive potential present on $\Sigma$ particles as can be
seen in Fig. 4.
We also notice in Fig.4 that the NLE2 EOS does not change significantly on 
the choice of repulsive $\Sigma$ potential in contrast to NLE1 EOS.
\par
The predicted maximum neutron star masses are very close to the 
observational values for NLE1 and NLE2 EOS.
The results for the neutron star structure parameters are tabulated in 
Table I and central density vs mass are plotted in Fig 5.
From Fig. 5 and Table I, we observe that the maximum masses of the stable
neutron stars are 2.18$M_{\odot}$, 1.94$M_{\odot}$ and 1.97$M_{\odot}$ and 
corresponding radii are 11.9$km$, 10.7$km$ and 10.8$km$ for NL3, NLE1 and 
NLE2 EOS respectively. 
The corresponding central densities are $2.0\times 10^{15}~ g~ cm^{-3}$, 
$2.2\times 10^{15}~ g~ cm^{-3}$ and $2.2\times 10^{15}~ g~ cm^{-3}$ (
$>$ 7 times nuclear matter density) for NL3, NLE1 and NLE2 respectively at 
the maximum neutron star masses.
These maximum masses calculate in our models are in the range of recent 
observations \cite{para98,barz99,oros99,mill98}, where the observational 
consequence are discussed below.
Very recently, it has been observed that the best determined neutron star 
masses \cite{thor99} are
found in binary pulsars and are all lie in the range 1.35$\pm 0.04M_{\odot}$
except for the non-relativistic pulsars PSR J1012+5307 of mass $M=(2.1\pm
0.8)M_{\odot}$ \cite{para98}.  
There are several X-ray binary masses have been measured, the heaviest
among them are Vela X-1 with $M= (1.9\pm 0.2)M_{\odot}$ \cite{barz99} 
and Cygnus X-2 with $M= (1.8\pm 0.4)M_{\odot}$ \cite{oros99}.
From recent discovery of high-frequency brightness oscillations in 
low-mass X- ray binaries, the large masses of the neutron 
star in $QPO 4U 1820-30 (M=2.3)M_{\odot}$ \cite{mill98} is confirmed and this
provides a new method to determine the masses and radii of the 
neutron stars.
We also tabulate the moment of inertia and the surface red shift in Table I, 
which are important for the dynamic and transport properties of pulsars.
\par
At this point, we argue that the softening of EOS may lead to kaon 
condensation in neutron stars \cite{li97} and hence may give a constraint 
on the best determined maximum mass \cite{thor99}.
However, we feel that from the KaoS data on kaon production, together with
kaon flow from heavy-ion reactions \cite{part95}, it is important to know
the momentum dependent $K^+$ and $K^-$ potentials in dense matter in 
contrast to the prediction of the chiral perturbation theory.
In the present calculation, we don't explore this, but work is in 
progress \cite{sahup2} by implementing the same momentum forces as given 
in eqs (2,3).
\section{Summary}
We have described the nuclear EOS in the frame work of relativistic mean
field theory using a relativistic transport model in the heavy-ion collisions.
From the heavy-ion collisions data, more specifically, the baryon flow for
$Au + Au$ systems at SIS to AGS energies and above we noticed that the strength
of the vector potential has to be reduced substantially at high density and
high relative momenta to describe the experimentally observed flow data at 
1-10 A GeV.
In a different way, the vector potential should be non- linear function 
of the baryon density.
We took this effect by introducing the momentum dependent cut-off into
the vector potential in contrast to the heavy-ion collision data. 
We use the same dynamic treatment in our relativistic mean field model to 
calculate the nuclear EOS.
It is found that the derived nuclear EOS is moderately soft at density
$\ge 7 \rho_0$ than the original considered nuclear EOS without momentum 
dependent potentials.
This due to the reduction of repulsive nuclear interaction in the nuclear
EOS at high density.
We then employ the same nuclear EOS to the neutron star structure calculation.
\par
In the neutron star matter, the core of the neutron stars are considered to
be composed of neutrons along with an admixture of protons, electrons,
muons and hyperons at zero temperature.
The resulting maximum mass of stable neutron stars are $2.18M_{\odot}$, 
$1.94M_{\odot}$ and $1.97M_{\odot}$ for the NL3, NLE1 and NLE2 models, 
respectively.
We observed that the maximum mass of the neutron
star for NLE1 and NLE2 are lower than that for NL3 due to a reduction 
of the vector field at higher densities.
Also, we noticed that the potential felt by $\Sigma$ particles is
not so relevant to neutron star structure calculation.
The corresponding neutron star radii are 11.9 $km$, 10.7$km$ and 10.8$km$
for NL3, NLE1 and NLE2, respectively, whereas the corresponding central densities
are $2.0\times 10^{15}~ g~ cm^{-3}$, $2.2\times 10^{15}~ g~ cm^{-3}$ and $2.2\times 
10^{15}~ g~ cm^{-3}$ respectively for NL3, NLE1 and NLE2 at the maximum neutron 
star mass. 
We found that the maximum mass for NLE1 and NLE2 are in the observable region 
\cite{para98,barz99,oros99,mill98}, $1.4M_{\odot} < M_{max} < 2.2M_{\odot}$
and the corresponding radius is in between 8-12 $km$.
\vskip 0.4 in
The author likes to thank W. Cassing and A. Ohnishi for critical
reading and W. Cassing, A. Ohnishi and Y. Akaishi for fruitful discussions. 
PKS likes to acknowledge the support from the Japan Society for the 
Promotion of Science, Japan. 
This work is dedicated to Prof. Bhaskar Datta, who passed away on 3rd 
November 1999.
\vskip 0.2 in
\begin{table}
\noindent {Table I}
\vskip 0.1 in
\begin{tabular}{ccccccccc}
\hline
\multicolumn{1}{c}{$\varepsilon_c$}&
\multicolumn{1}{c}{$R$} &
\multicolumn{1}{c}{$M/M_{\odot}$} &
\multicolumn{1}{c}{$z$} &
\multicolumn{1}{c}{$I$} &
\multicolumn{1}{c}{} \\
\multicolumn{1}{c}{($g~cm^{-3}$) } &
\multicolumn{1}{c}{($km$)} &
\multicolumn{1}{c}{} &
\multicolumn{1}{c}{}&
\multicolumn{1}{c}{($g~cm^2$)} &
\multicolumn{1}{c}{}\\
\hline
6.0 E14&10.82&1.00&0.17&9.55 E44& \\
8.0 E14&11.46&1.44&0.26&1.61 E45& \\
1.0 E15&11.54&1.66&0.32&1.94 E45& \\
1.5 E15&11.27&1.89&0.41&2.15 E45& \\
2.0 E15&10.89&1.94&0.45&2.08 E45& NLE1 \\
2.5 E15&10.55&1.94&0.48&1.96 E45& \\
3.0 E15&10.26&1.93&0.50&1.83 E45& \\
4.0 E15&9.83&1.88&0.51&1.62 E45& \\
\hline
6.0 E14&10.87&1.03&0.18&9.93 E44& \\
8.0 E14&11.55&1.48&0.27&1.70 E45& \\
1.0 E15&11.63&1.72&0.33&2.05 E45& \\
1.5 E15&11.36&1.92&0.41&2.24 E45& \\
2.0 E15&10.98&1.97&0.46&2.15 E45& NLE2 \\
2.5 E15&10.96&1.97&0.47&2.02 E45& \\
3.0 E15&10.37&1.95&0.50&1.88 E45& \\
4.0 E15&9.92&1.89&0.51&1.66 E45& \\
\hline
6.0 E14&13.20&1.60&0.25&2.31 E45& \\
8.0 E14&13.26&1.89&0.31&2.84 E45& \\
1.0 E15&13.08&2.04&0.36&3.02 E45& \\
1.5 E15&12.43&2.17&0.44&2.96 E45& \\
2.0 E15&11.85&2.18&0.48&2.72 E45& NL3 \\
2.5 E15&11.38&2.16&0.51&2.49 E45& \\
3.0 E15&11.00&2.13&0.53&2.28 E45& \\
4.0 E15&10.40&2.05&0.55&1.95 E45& \\
\hline
\end{tabular}
\end{table}

\newpage
\vskip 0.1cm
{{\psfig{figure=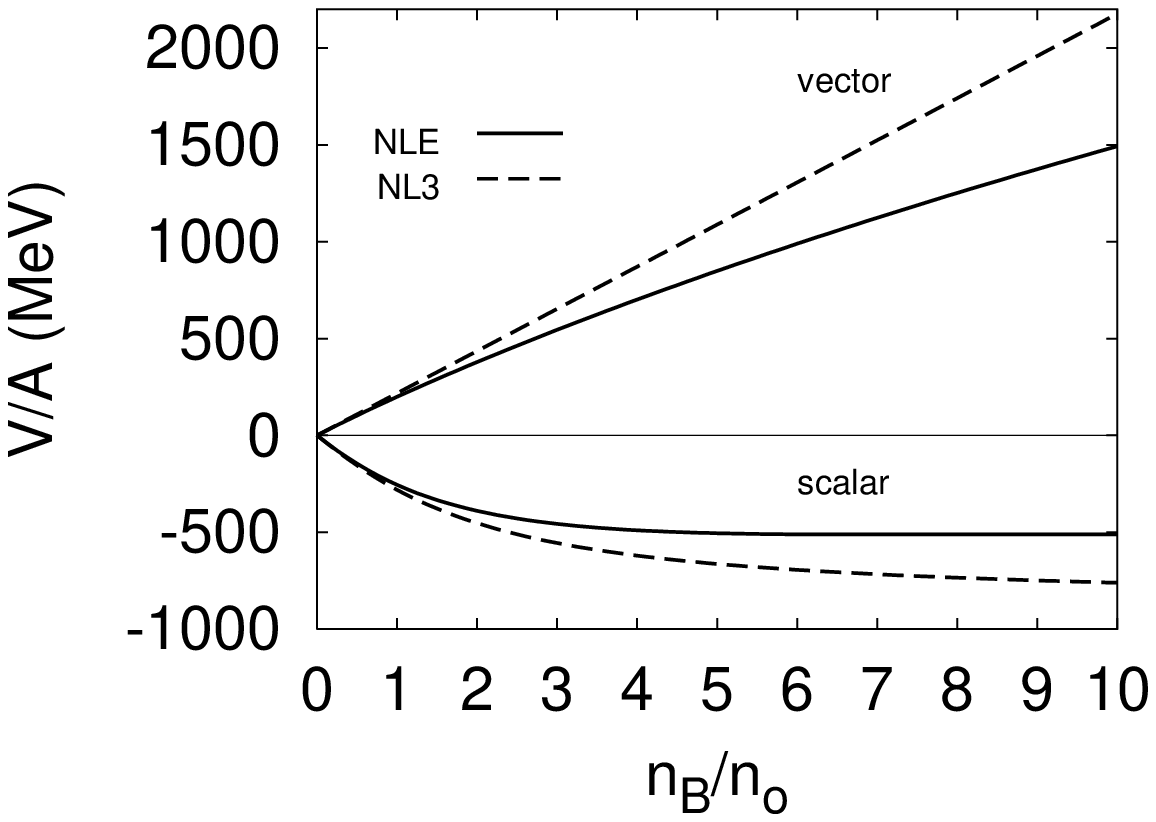,height=12cm,width=18cm}}
{\center {\bf Fig.1 potential energy per nucleon vs baryon density in 
units of $n_0$. The solid lines (NLE) are momentum dependent
potentials and the dashed lines (NL3) are without momentum dependent
potentials (see the text).}} 
\vskip 0.1cm
{{\psfig{figure=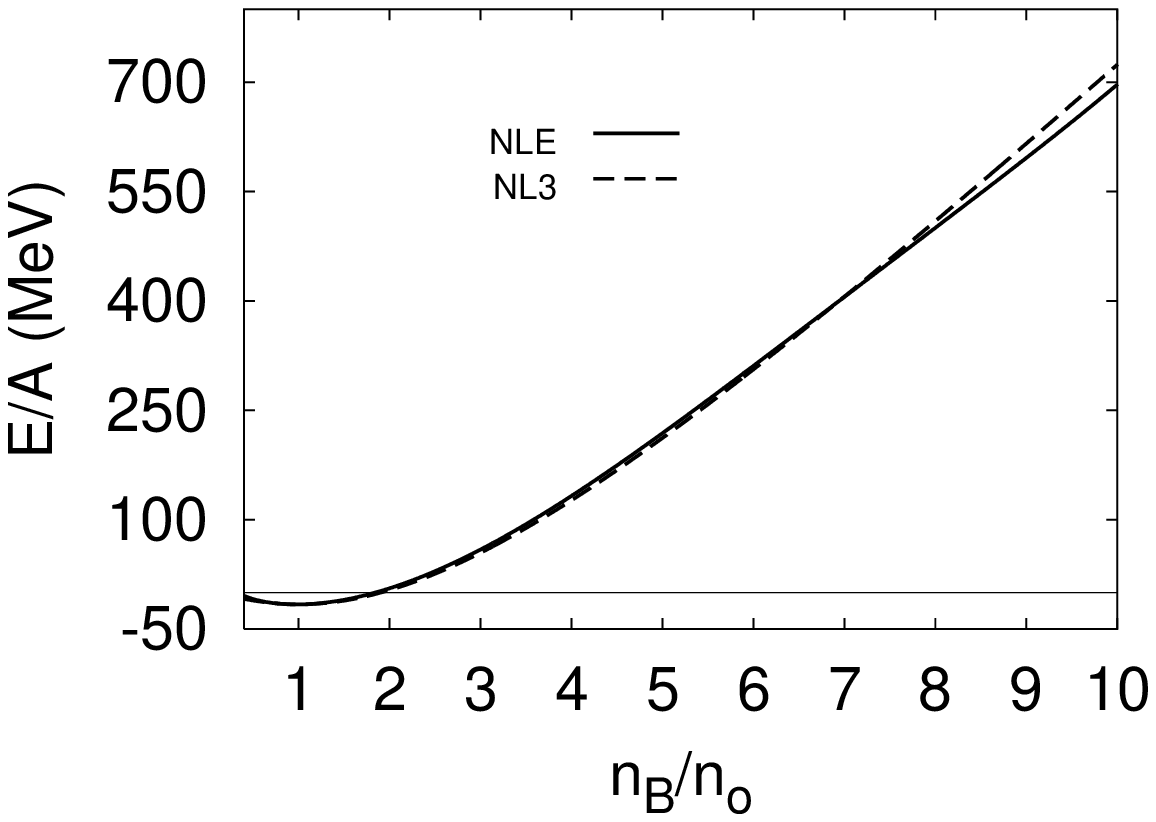,height=12cm,width=18cm}}
{\center {\bf Fig.2 Energy per nucleon vs baryon density in units of $n_0$. 
The models are same as Fig. 1.}} 
\vskip 0.1 cm
{{\psfig{figure=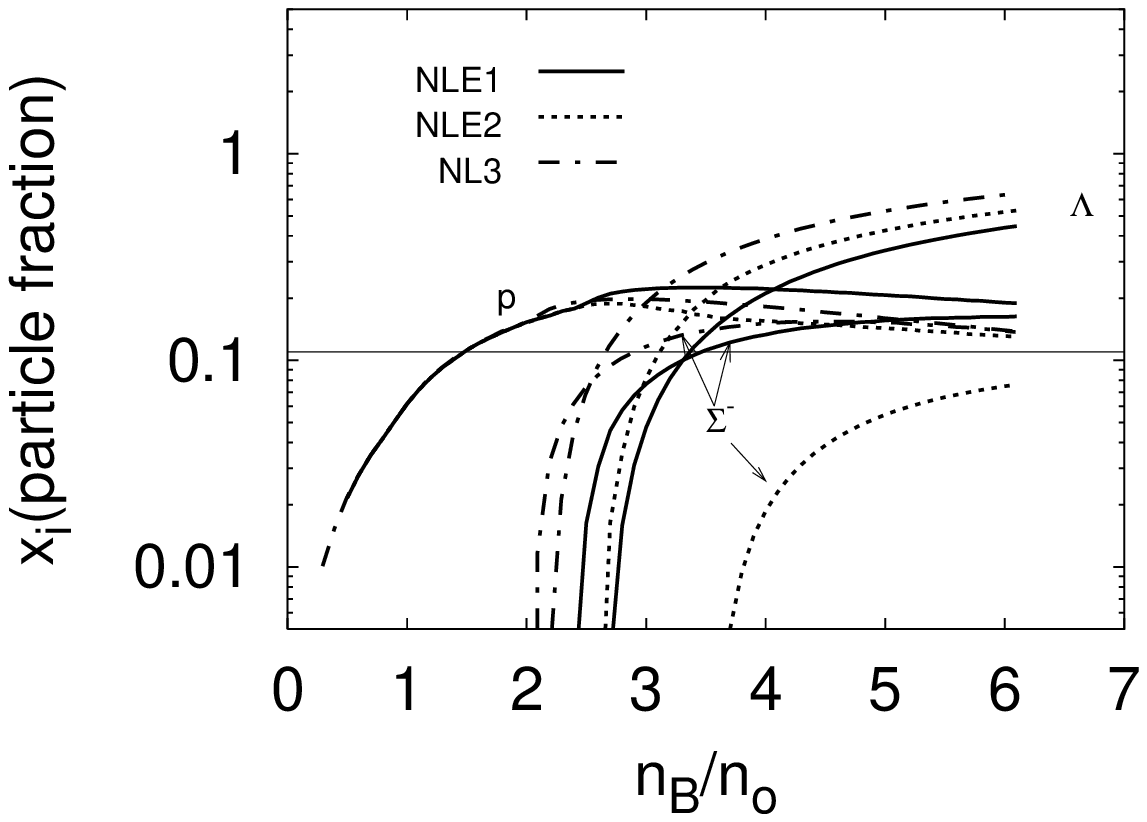,height=12cm,width=18cm}}
{\center {\bf Fig.3 The concentration of each particle ($x_i=n_i/n_B$) vs 
baryon density in units of $n_0$. The momentum dependent potentials have been
incorporated in NLE1 (solid line) and NLE2 (dashed line). The
dashed-dot lines (NL3) are without momentum dependent potentials.
The potential are seen by $\Lambda$ and $\Sigma$ are same in NLE1, 
NL3 (dashed-dot line) and are different in NLE2 (see text).
}}
\vskip 0.1 cm
{{\psfig{figure=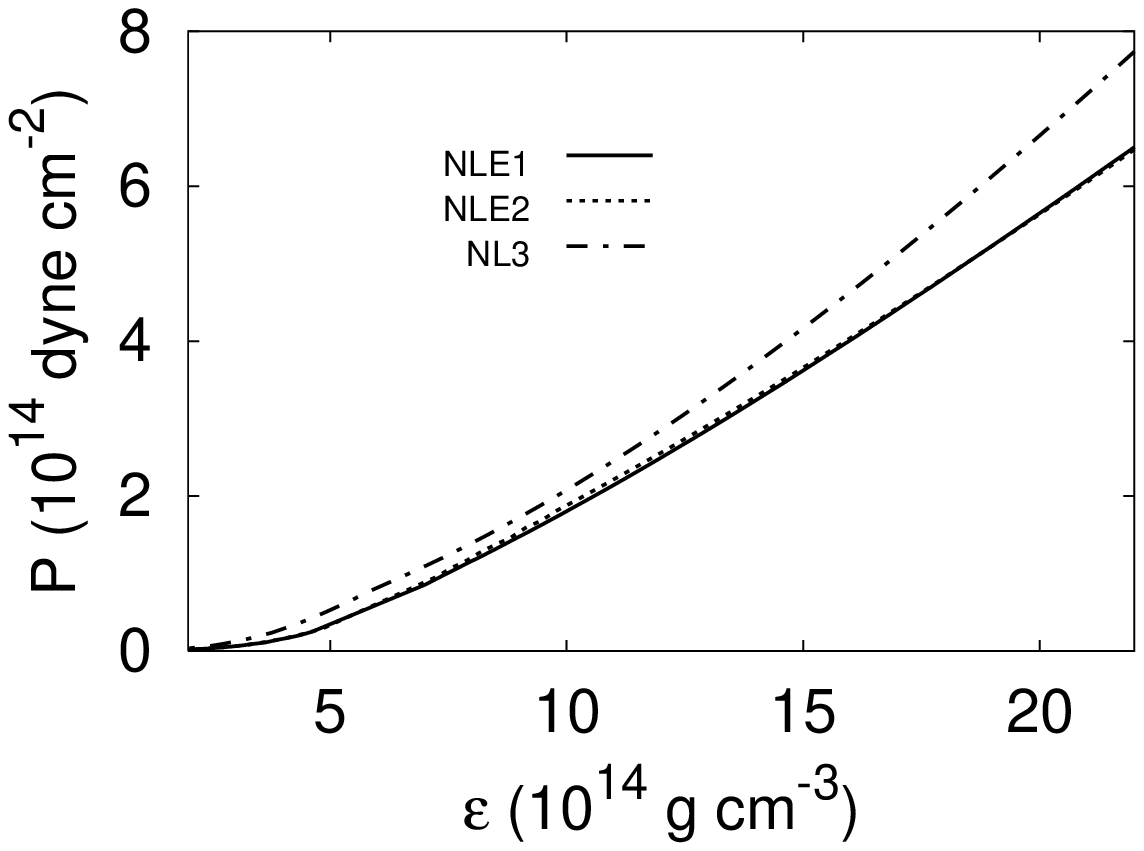,height=12cm,width=18cm}} 
{\center {\bf Fig.4 The neutron star matter pressure vs the energy density.
The models are same as Fig. 3. }} 
\vskip 0.1 cm
{{\psfig{figure=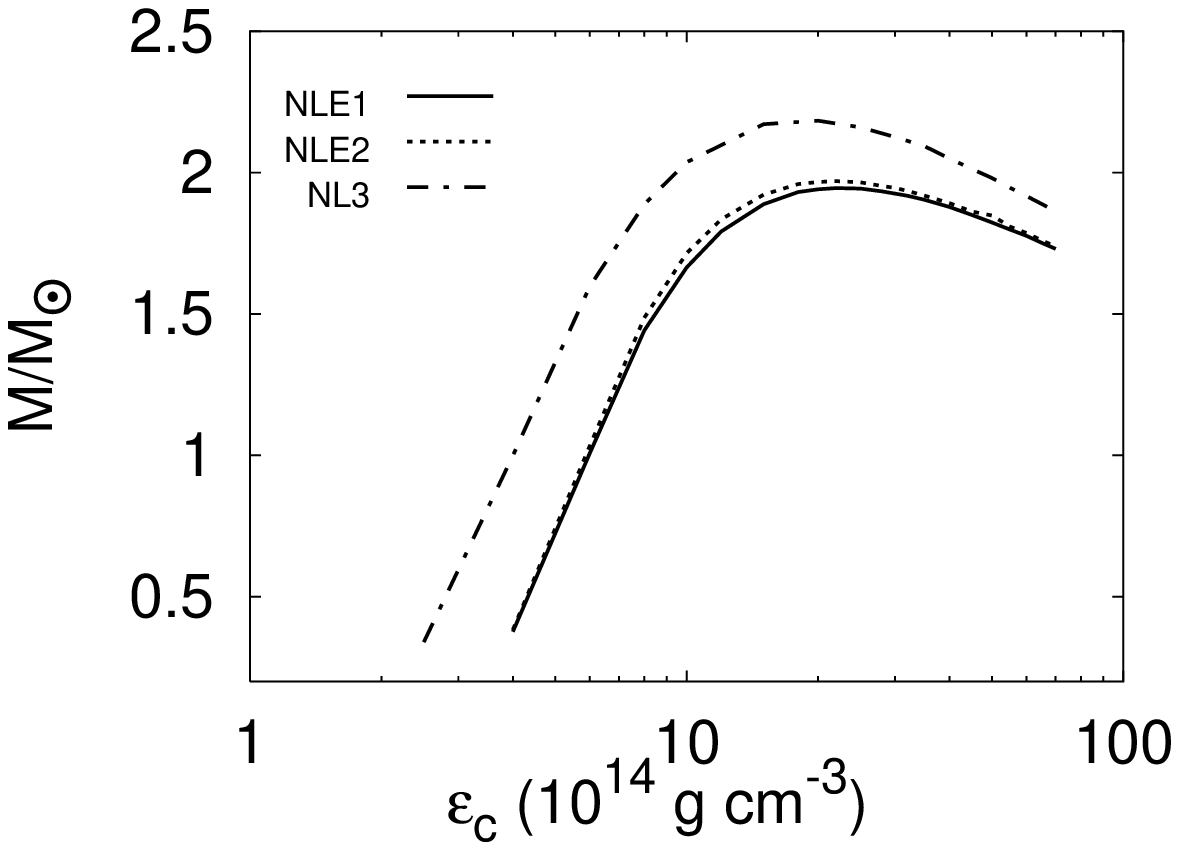,height=12cm,width=18cm}} 
{\center {\bf Fig.5 The neutron star mass vs radius. The models are
same as Fig. 3.}} 
\end{document}